\documentstyle[11pt,epsfig]{article}
\begin{document}
\title{Large Angle Beamstrahlung as a Beam-beam 
Monitoring Tool}
\author{G. Bonvicini\\
{\it Wayne State University, Detroit MI 48201}
\and
J. Welch\\
{\it Cornell University, Ithaca NY 14853}} 
\date{\today}
\maketitle

\section{Introduction}

\begin{figure}
 \begin{center}
    \mbox{\epsffile{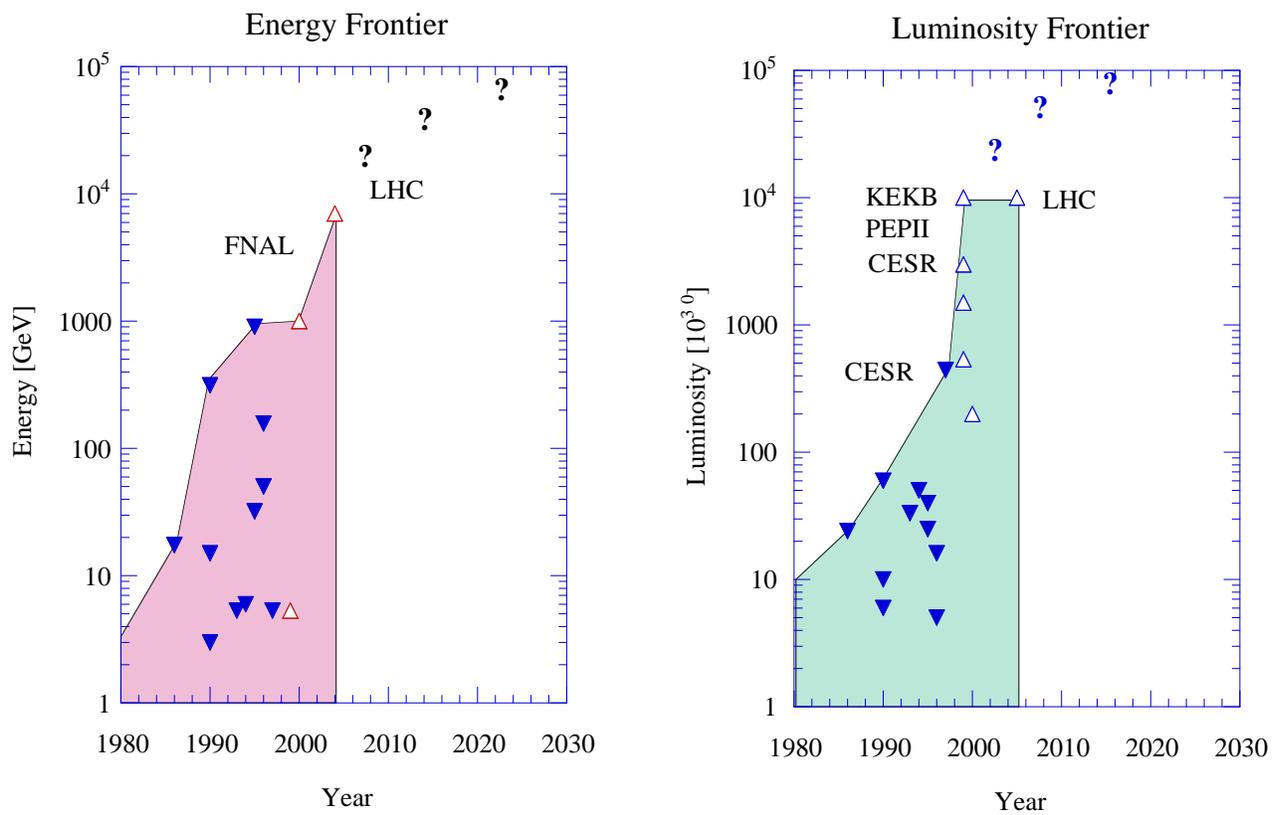}}
    \caption{ Solid point represent actual obtained peak 
luminosities (right
plot) and energies (left plot) in selected high energy 
physics accelerators.
Open points represent planned values for machines under 
construction.
\label{fn:Livingston} }     
 \end{center}
\end{figure}

Luminosity in high energy physics colliders is now more  important than ever
before. A comparison of `Livingston' charts,  figure~\ref{fn:Livingston}, shows
the machine luminosity increasing at a rate faster than the increase in 
machine energy. Now and in the  forseeable future extremely high luminosity is
needed to produce meaningful  quantities  of rare events (factories) and to
cope with the $1/s$ dependence of the  production cross section of high mass
states.

Beam alignment complexities with two independent rings,  dynamic beta effects,
disruption, crossing angle collisions and beamstrahlung  are just some of the
new phenomena to be dealt with in modern machines  running at peak luminosity.
Additionally, high luminosity machines must operate   near design luminosities
in order to be useful, so there is a strong motivation  for understanding,
controlling, and possibly taking advantage of some these  new high luminosity
phenomena that occur at the interaction point.

Luminosity for two gaussian beams colliding head-on with  equal beam sizes is,
\begin{equation}
L	= {f N_1 N_2 \over 4\pi \sigma_x \sigma_y}, 
\label{en:luminosity}
\end{equation}
where $f$ is the bunch collision frequency, $N_{1(2)}$ is  number of particles
per bunch in beam 1(2), and $\sigma_{x(y)}$ is the  horizontal(vertical) beam
size. Beams are usually quite flat, ($r=\sigma_y/\sigma_x<<1$), with $\sigma_y$
being typically a few microns or less. The two beam axes must be continously
aligned to  intersect to much better than $\sigma_y$ or luminosity is  lost. 

For high luminosity conditions, the charge per unit area is made so large that 
the interaction of one beam with the electromagnetic fields of
the other beam can cause poor lifetime, beam blow-up,  and other types of
instabilities.

One effect of the interaction is to cause  the beam size to change
significantly over a distance comparable to  the length of the bunch. This is
called  disruption. Since changing the beam size also changes the strength of
the interaction, disruption is a very complicated, non-linear phenomenon.

If the beam-beam deflections are significant compared with  the average angular
spread of particles in the bunch, then the target beam   acts like a short high
gradient magnet and, in a storage ring,  distorts the beta  functions all
around the ring. This effect is called the dynamic beta effect and has been 
observed directly at CESR \cite{cinabro}. It is also a highly non-linear effect
because the  distortion seen by small amplitude particles is largest.

In this paper we propose a method, which has never been  
used before, to
monitor the beam-beam collision conditions. The  
strength of this method,
besides its low cost, versatility, and strong signals 
well  separated in angle
from machine backgrounds, is the  direct visualization 
of many aspects of the
beam-beam  interaction.   

As in many other areas of physics research, 
visualization of a complex
phenomenon is at a premium. In this case, we believe a 
beamstrahlung detector
looking at the interaction point might help in faster 
and better diagnosis of
misalignments and incongruities of beam sizes during 
collision. It might also
shed some light on many of the of the dynamical effects 
which limit the maximum
tuneshift parameter and hence the luminosity for a given 
current. Considering
the difficulty in getting a machine into top performance 
and keeping it there,
an instrument, such as we propose,  which can directly 
observe the beam-beam
interaction  could prove to be quite valuable.

We predict that, if the first such monitor is 
successfully operated, it will be
adopted  by the three $e^+e^-$ B-factories (CESR, PEP 
II,  and KEK B), and by
the $\phi$-factory in Frascati, and  rapidly become the 
primary beam-beam
monitor at all facilities, just as a similar  device 
became primary beam-beam
monitor at the SLC. The use of such a monitor by HERA 
(proton side only) is
also a  possibility.

We start by discussing the physics of beamstrahlung and 
what
might be expected at the various high luminosity 
factories and the SLC. Then we
go into some detail about `short magnet' radiation and 
what it can tell about
the collision. Detector design, backgrounds, and 
possible implementation at
CESR are then discussed. Finally we address some 
`frequently asked questions'
concerning beamstrahlung.

\section{Beamstrahlung}
Beamstrahlung occurs when  particles of the first beam, 
or radiating beam, are
deflected by the  electromagnetic field of the second 
(target) beam and emit
synchrotron radiation. Beamstrahlung was  observed first 
at the
SLC \cite{bonvi}.

As long as the radiated energy is small compared to the  
beam energy, most but
not all the properties of beamstrahlung can be derived  
using the  formulae in
Jackson's ``Classical Electrodynamics'' \cite{jacks1},  
convoluted over the
space-time distribution of charges,  and applying usual 
ultrarelativistic
approximations. The important  paper \cite{albert} 
calculates a number of
low-energy beamstrahlung  properties, which are very 
relevant to this paper.

Under the conditions of Eq.~\ref{en:luminosity}, assuming 
beams of  equal mass
and  energy,  the beamstrahlung power $W_1$ 
 is
\begin{eqnarray}
W_1 	&=& fU_1 \\
	&=& g(r)  r_e^3 m c^2 \gamma^2 
		{f N_1 N_2^2\over  
\sigma_x\sigma_y\sigma_z}
		\label{en:Power} \\
	&=& 4 \pi g(r) r_e^3 mc^2 \gamma^2 {L N_2\over  
\sigma_z},
\end{eqnarray}
where $U_1$ is the energy radiated per collision \cite{albert}.
Here $r_e$ is the classical radius of the beam  particles, $m$ their mass, and
$\gamma=E/m$ the relativistic factor. $g(r)$ is a  dimensionless factor 
obtained in the integration over space-time, equal to\cite{albert}
\[ g(r) = {64\sqrt{\pi}r\over 3\sqrt{3r^4-10r^2+3}}\arctan{({
\sqrt{3r^4-10r^2+3}\over 3r^2+8r+3})}.\]
It is maximal for round beams ($r=1$), at 2.735..., and for flat beams ($r$
small) it can be approximated as follows
\[ g(r) \sim 11.4 r.\]
In the flat beam limit, $\sigma_y$ cancels in  Eq.~\ref{en:Power} and the
dependence of $W$ on beam parameters becomes
\begin{eqnarray}
W\propto {\gamma^2 N_1N_2^2\over 
\sigma_z\sigma_x^2}.
\end{eqnarray}
All of these parameters are well  known by other means. There is little
information in the total emitted  power, all of the non-trivial information is
in the asymmetries  described below. On the other hand, total power becomes a
reliably  predicted quantity that we use as a constant in the following
calculations.

Table~\ref{tn:Machines} summarizes the beam parameters 
at various  facilities,
according to Ref.~\cite{PDB} and their  beamstrahlung 
power at design
luminosity. It is  surprising to discover that 
beamstrahlung yields at the SLC,
PEP-II and KEK are  roughly the same, with the higher 
luminosity compensating
for the lower  energy and longer beams.
\begin{table}
\caption{Beamstrahlung parameters derived for high 
luminosity factories and
SLC.\label{tn:Machines}}
\begin{center}
\begin{tabular}{|l|l|r|r|r|r|r|}
\hline
		&Machine& SLC 	& CESR 1997
                		& PEP	& KEK	
& Frascati\\
\hline
$L_{max}$ 	&$[10^{33} cm^{-2}s^{-1}]$ 
			&0.0008	& 0.44 	& 3. 	& 10. 	
& 0.13 \\
$r$ 		&	& 0.3 	& 0.022	& 0.04 	& 0.025 	
& 0.01 \\  
$\sigma_z$	&$[mm]$ & 0.8 	& 18 	& 10 	& 4 	
& 30 \\   
$E_1$		&$[GeV]$& 50 	& 5.3 	& 3.1 	& 3.5 	
& 0.51 \\  
$E_2$		&$[GeV]$&-	& - 	& 9 	& 8 	
& - \\
$N_1$ 		&$[10^{10}]$ 	
			&3.5	& 14  	& 5.9 	& 3.2 	
& 9\\  
$N_2$ 		&$[10^{10}]$ 	
			&-	& - 	& 2.7 	& 1.3 	
& -\\
$B_{\sigma 1}$ 	&$[T]$ 	& 200 	& 0.14 	& 0.17 	& 0.43  	
& 0.01 \\
$B_{\sigma 2}$ 	&$[T]$ 	& - 	& - 	& 0.33 	& 1.0  	
& - \\
$W_1$ 		&$[W]$	& 12 	& 1.6 	& 2.3 	& 7.5  	
& 0.0008\\  
$W_2$		&$[W]$	& -	& - 	& 38 	& 95 	
& - \\
$\theta_o$	&$[mrad]$ 	
			& 27  	& 6 	& 7 	& 12 	
& 5 \\
$\nu_1$		&$[10^{10}s^{-1}]$	
			& $7\times 10^{-7}$ 
				& 0.34 	& 2.0 	& 0.32 	
& 4.6 \\
$\nu_2$		&$[10^{10}s^{-1}]$	
			&- 	& - 	& 1.2 	& 0.15	
& - \\
\hline
\end{tabular}
\end{center}
\end{table}
Two of the three B-factories feature beams crossing at 
an  horizontal  angle,
and two  feature unequal energy beams. The crossing 
angle  $\alpha_x$ is
respectively  2 and 10 mrad at CESR and KEK, and results 
in changes in  the
total radiated power of $O(\alpha_x^2)$. The crossing 
angle can be  neglected
except for a trivial distortion of the small angle 
distributions.

The effect of unequal beams on Eqs.~\ref{en:Power} is 
readily derived with a
simple substitution. $\gamma$ is the relativistic factor 
of the radiating beam,
and $\sigma_z$ the length of the target beam, and all 
quantities are defined in
the laboratory  frame. With these substitutions, 
Eqs.~\ref{en:Power} are valid
for beams of unequal energies and beam lengths,  such as 
at PEP-II or KEK, in
the limit of rigid beams.

Current CESR conditions have to be considered the  
benchmark.
Table~\ref{tn:Machines} lists the total radiated power  
$W$, as well as other
meaningful quantities. At present day CESR, one obtains 
a total beamstrahlung
power of 2.4 W. At PEP-II, the radiated beamstrahlung 
power is respectively 4
and 80 W for the low and high energy beams.

Beamstrahlung has been the subject of many theoretical 
calculations,
which are not relevant to this paper,
and is a crucial issue for future linear colliders. At 
the SLC it was
detected and monitored primarily to monitor the 
collision of two beams
coming from different beamlines, and therefore having 
different transverse
dimensions and even orientations in the transverse 
plane. 
There were 
seven degrees of freedom in the transverse plane at the 
SLC
(the four beam dimensions,
the angle between the two major axes of the transverse 
ellipses, and the two
coordinates of the beam-beam offset).
We notice that
the situation at asymmetric B-factories is quite similar 
to the SLC - 
two beams totally independent of each other. It is at 
those machines
that the device described here will be most useful. 
The situation at CESR is somewhat
more constrained by the single-ring machine.

The North and South beamstrahlung detectors described in 
Refs.~\cite{bonvi} and 
\cite{bonvi1} lasted 6 and 8 years at the SLC  without 
particular problems,
delivering instant and  primary information on beam-beam 
overlap to the
operators, via a monitor placed directly  above the 
operator's desk. During
their lifetime they absorbed  close to 100 GRad$/$year  
and were designed to
minimize visible radiation  backgrounds, a fact relevant 
to this paper.

The SLC monitors worked because the effective magnetic  
field accompanying the
beams, at a typical distance of one standard deviation 
from the beam center,
is 
\begin{equation}
B_\sigma\sim  {2 N_2 r_e m c\over e\sigma_z\sigma_x}\\
\end{equation}
which is now of order 100 T (and was of order 10 T when  
beamstrahlung  was
first observed \cite{bonvi}). 

10 Tesla far exceeds all  other magnetic fields
along the beam line. Thus the beamstrahlung critical  
energy was much higher
than the beam line synchrotron radiation, and  although 
the  synchrotron
radiation power was one million times higher  at the 
time of first observation,
a detector with favorable signal/noise could be built.

Table~\ref{tn:Machines} shows that $B_{\sigma}$ is quite 
low at today's 
B-factories,  and comparable or lower than the beam line 
magnetic fields. A
signal in this case is extracted by noticing that the 
beam,  considered as a
magnet, is much shorter than the other magnets. Thus the 
Fourier  transform 
produces a drastically different radiation spectrum.

\begin{figure}
 \begin{center}
    \mbox{\epsffile{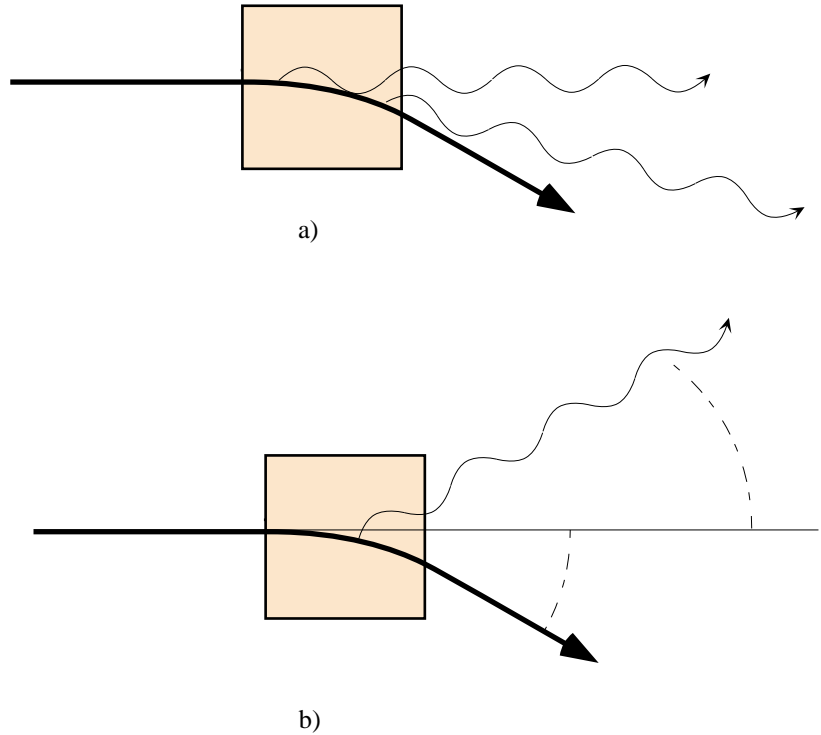}}
    \caption
        {    
   Searchlight approximation: the observation angle
   is smaller than the bending angle. b) Short magnet 
   approximation: the 
   observation angle is larger than the bending angle.
     \label{fig:SEARCHLIGHT}
     }
 \end{center}
\end{figure}

``Short magnet'' (SM) 
radiation replaces the usual synchrotron radiation 
spectrum 
when the observation angle (Fig.~\ref{fig:SEARCHLIGHT}) 
is much larger than 
the deflection angle.
SM extra terms contribute a fraction of the power of 
order
$1/\gamma^2$ of the total power (that is, they  
integrate to a quantity
that has no power dependence on $\gamma$), 
and therefore are perfectly consistent with
the ``searchlight'' approximations for synchrotron 
radiation \cite{jacks1}.
The use of SM radiation as a beam monitor 
was first suggested in 
Ref.~\cite{albert}. A practical detector was 
suggested first at the SLC in Ref.~\cite{bonvi3}. 

\section{Properties of Short Magnet Radiation}
\begin{figure}
 \begin{center}
    \mbox{\epsffile{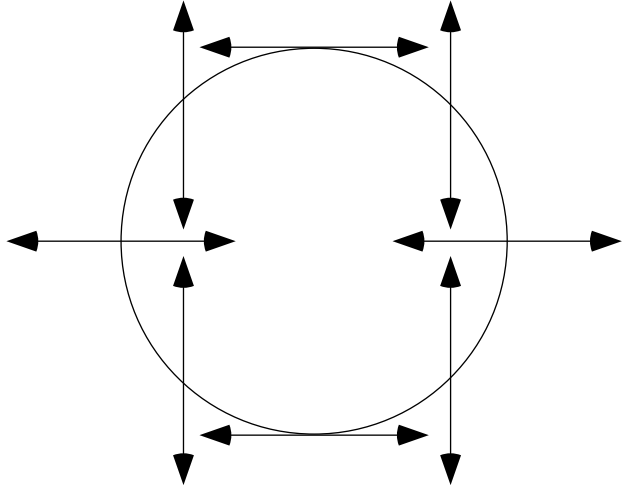}}
    \caption[]
        {    
Eightfold polarization pattern for a particle bent in 
the
horizontal direction.
\label{fig:EIGHTFOLD}
        }
 \end{center}
\end{figure}

The properties of SM radiation were first investigated 
by 
Coisson \cite{coisso}. SM radiation was first observed 
in
Ref.~\cite{alber1}.
In regions where SM radiation dominates,
such as at large angle ($\gamma\theta>>1$), which are 
the only ones
relevant to this paper, three properties are of 
interest:
\begin{itemize}
\item The large angle power is proportional 
to $1/\gamma^2$ (or a fraction of $1/\gamma^4$ of the total power) . 
One of the main points of this paper is to show that a 
possible, 
but somewhat marginal,
signal at the SLC \cite{bonvi3} becomes large at the 
lower energy 
factories. 
\item the radiation spectrum is gaussian, and the cutoff 
frequency
independent of beam energy. The spectrum extends far 
beyond the
spectrum calculated in a standard way.
\item radiation is linearly polarized parallel or 
perpendicular
to the acceleration, with an 
eight-fold symmetry in azimuth.
Fig.~\ref{fig:EIGHTFOLD} shows the polarization pattern 
for a vertical 
bend.
\end{itemize}

The large-angle power emitted as a function of angle
and frequency is expressed as \cite{albert}
\begin{equation}
 I(\theta,\phi,\omega)=
 {3\sigma_z W_1\over 4\pi\sqrt{\pi} 
c}{1\over\gamma^4\theta^3} 
\exp{({-\sigma_z^2\theta^4\omega^2\over 16c^2})}. 
\label{en:LargeAngle}
\end{equation}
For reasons that will become clear in the next Section, 
we compute the
power in the visible
($3\times10^{15}<\omega<6\times10^{15}s^{-1}$), that 
falls
within 6\% of the azimuth, and one mrad upward of 
the optimal angle discussed below (from 6 to 7 mrad for
CESR, from 8 to 9 mrad for PEP-II). This is the power 
readily available
for a simple beamstrahlung monitor looking at the 
beam-beam interaction
by means of a few small viewports drilled through the 
beam pipe. 
At the optimal angle the exponential
factor in Eq.~6 is equal to -2 for photons at the lower 
$\omega$ limit.
For the purposes of this integration the frequency upper 
limit
can be taken to infinity, the error function 
at -2 is equal to 0.025 and the simple final result is
\begin{equation}
W_{exp}= {0.0225 W_1 \over 
\gamma^4}({1\over\theta_{min}^4}-
{1\over\theta_{max}^4}).
\end{equation}
The total beamstrahlung power in the narrow phase space 
region defined
above is $0.8\times 10^{-9}$ of the total power for 
CESR. 
When divided by the average
photon energy (2 eV) one gets the number of photons per 
second $\nu_1$,
listed in Table~\ref{tn:Machines}.
Current CESR conditions will deliver some 3.5 billion
visible photons
per second in the region under consideration, or about
1000 photons per beam-beam collision.
Clearly this method will give
plenty of signal even when
the luminosity is one or two orders of magnitude lower 
than design.

In the case of beam-beam offsets, and general beam-beam  
diagnostics, things
can become quite complex.  A good treatise of 
beamstrahlung patterns versus
beam conditions can be found in Gero's  thesis 
\cite{gero}. A desired result for
a particular beam-beam  configuration can always be 
obtained with a
straightforward numerical integration  over  space-time. 
A detailed treatment
of all possible  pathologies  goes far beyond the scope 
of a short paper, so we
will  limit ourselves to a few comments, plus the 
concepts outlined in the last
Section to give an idea of the power of the method.

\begin{figure}
 \begin{center}
    \mbox{\epsfxsize12.5cm\epsffile{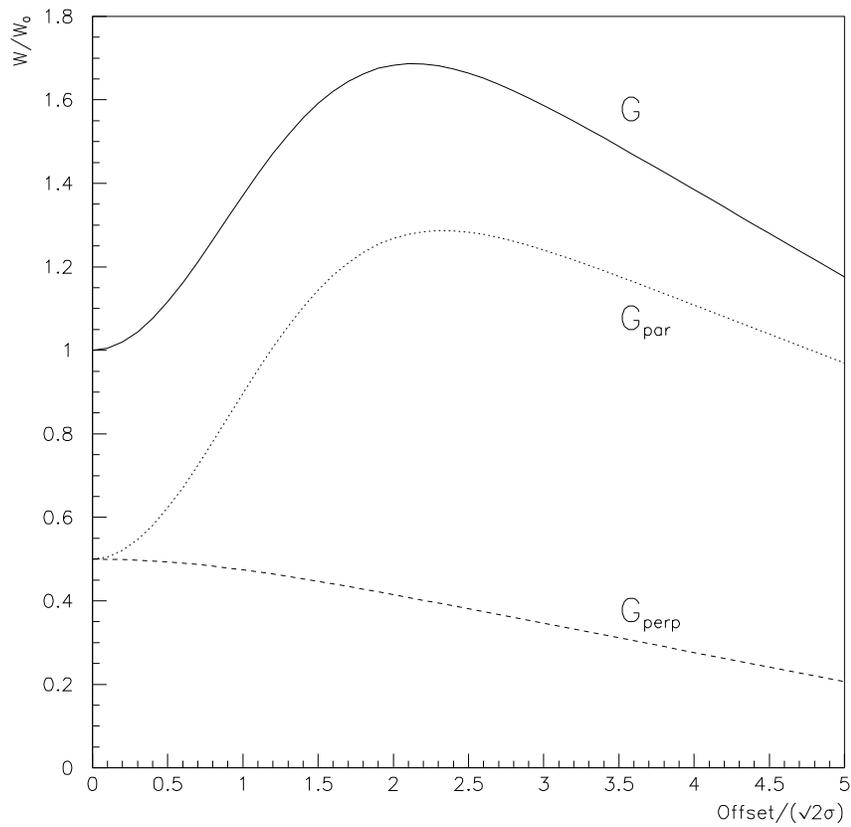}}
  \caption [ ]
          {
The total power and the components polarized
respectively parallel and perpendicular to the bending 
plane, as a function
of the normalized offset.
  \label{fig:FORMAF}
          }
 \end{center}
\end{figure}

A beam-beam offset will necessarily generate a dipole  
moment, which will be
reflected in a non-zero polarization of the emitted  
light, as well as a change
in the total light yield. We recall only formulae for 
the important case of
flat  beams, experiencing an offset along $y$ 
\cite{albert}. Dividing the power 
emitted into a component parallel to the offset and 
perpendicular to the
offset,  the result is
\begin{eqnarray}
I_\perp(\theta,\phi,\omega)&=& I(\theta,\phi,\omega)
{G_\perp \cos^2(2\phi)+G_\parallel \sin^2(2\phi)\over 
G},\\
I_\parallel(\theta,\phi,\omega)&=& I(\theta,\phi,\omega)
{G_\perp \sin^2(2\phi)+G_\parallel \cos^2(2\phi)\over 
G}.
\end{eqnarray}
The $G$ form factors are defined as
\[ G_\perp +G_\parallel = G(r,v),\;\;\; G(r,0)=g(r)/2.\]
The variable $v$ represents the offset in units of 
$\sqrt{2}\sigma_y$.
The polarization asymmetry is defined as
\[ A= {G_\perp -G_\parallel \over G(r,v)}.\]
Fig.~\ref{fig:FORMAF} shows the
$G$ form factors as a function of the offset $v$. The 
camelback shape is a
classical feature of beamstrahlung, due to the radiating 
beam sampling a
region of higher field, and has been used in the past to 
distinguish
beamstrahlung from the background \cite{bonvi}.

A lot of information about beamstrahlung patterns can be 
summarized as follows:
the radiation will have the same multipole pattern as 
the beams have in
the transverse plane. If beams are 
centred and have the
same axis, but have different aspect ratios, there will 
be also a quadrupole
pattern. A sextupole moment will develop if one of the 
beams is rotated
with respect to the other.

\section{Design of a Short Magnet Detector.}
The detector must consist of one or more transparent 
viewports in
the beam pipe, at a large enough angle so that 
synchrotron visible
light at zero degrees can be efficiently masked. The 
photons are then
separated by polarization and color and counted.

In the harsh environment surrounding a beamline 
photomultipliers,
with a bialkali photocathode, are known to work well 
\cite{bonvi1}.
Thus the frequency acceptance is fixed by the two 
frequencies where
the bialkali efficiency is at half-max ($3\times 
10^{15}<\omega<
6\times 10^{15}$ ). The lower limit is loosely called 
``red'' light and
the upper limit ``blue'' light.

We add an ``optimal'' condition for detection, that is 
that 
the exponent
in Eq.~6 for red light be equal to two, {\it ie}
\[ \theta_o = 2^{5/4} \sqrt{c\over \omega\sigma_z} = { 24\ 
mrad\over  \sqrt{\sigma_z(mm)}}.\]
The reason for such a choice is that all backgrounds
(Cherenkov 
radiation, direct synchrotron radiation, fluorescence in 
materials 
both in the beam pipe and in the detector) are constant 
within 60\% 
across the visible spectrum. By 
choosing red, and
setting the detector at an angle where the blue light 
is suppressed, we establish that a 
measure of blue light measures
the backgrounds. In other words,
\begin{eqnarray}
W_{red} &=& e^{-2}W_{signal} + W_{background}\\
W_{blue}&=& e^{-8}W_{signal}  + W
_{background}
\end{eqnarray}
The photon counter would consist of a prism splitting 
the
red and blue components. The blue light goes to its own 
photomultiplier.
The red light is separated in 
the horizontal and vertical polarization component. Each 
then goes
into a photomultiplier. This background subtraction 
scheme
would improve the sensitivity to a beamstrahlung signal 
by two
orders of magnitude. The exponential factor
in Eq.~\ref{en:LargeAngle} scales with $\omega\theta^2$. 
If background conditions demand a larger angle, the 
trick can still be
used by using a photocathode sensitive in the near 
infrared.

Timing information could be had with 0.67 GHz FADCs 
connected to
the PMTs, for the purpose of timing out sources of 
backgrounds
not coming from the interaction region. Timing would 
improve
the sensitivity to a beamstrahlung signal, in the 
presence of opposite
beam backgrounds, by one order of magnitude.

Ideally, a viewport should be located at a fixed grid in
azimuth, but avoiding
the eightfold pattern of Fig.~\ref{fig:EIGHTFOLD} (every 
45 degrees the same
information is replicated). Three viewports are needed 
to fully disentangle
the information, and
a viewport every 30 degrees is a possibility. 
In practice, it is best 
to stay out of the
synchrotron radiation sweep (the $x-$axis in Fig. 3), 
to avoid premature ageing of the window and quadrupole 
synchrotron 
radiation,
and monitor six other locations, for example at 30, 60 
and
90 degrees, and their opposite locations in azimuth. The 
radiation 
pattern is symmetric under
180 degrees rotations in azimuth, and half the 
monitors are redundant
for perfect beam alignment. The redundancy can be used 
to correct beam
misalignments. Twelve data would be obtained for seven  
possible degrees of
freedom.

The cons of this proposal are, obviously, beam pipe 
surgery, and increased
heat loads and RF leaks at the viewport location. The 
latter two can be
minimized by drilling several smaller holes per 
viewport. The heat load and
RF leaks scale like $1/b^4$, $b$ being the  hole 
radius, thus
replacing one hole with three would suppress the leaks 
by an order
of magnitude.
\section{Application at CESR \label{sn:Application}}

A beamstrahlung monitor is being considered for 
installation in the CESR
storage ring to aid in understanding beam-beam effects 
that occur at high
luminosity. CESR has both 
large dynamic beta effects as well as sizable 
disruption.

Horizontal and vertical disruption parameters are 
defined as the bunch length
divided by the nominal thin lens 
focal length in the plane of interest. For gaussian
bunches they work out to 
\begin{eqnarray}
D_{h/v} &=& 4\pi \xi_{h/v}/\sigma_z
\end{eqnarray}
In CESR they are often as high as 0.5 or more, which 
indicates that
there is substantial change in the transverse beam size 
during  collisions.

The dynamic beta effect is a distortion of beta 
functions seen by low amplitude
particles due to the beam-beam interaction. It is 
especially significant at high
tune shift parameter and when the tune is near a half 
integer.
For example theoretical values for 
$\beta^\ast_h /\beta^\ast_{h0} < .37$ for a horizontal 
tuneshift 
parameter of 0.032 and
horizontal betatron number of 0.526 \cite{Sagan}.

These types of 
nonlinear beam-beam distortions affect lowest amplitude 
particles the most;
beam tails are not affected as much.
But the low amplitude particles are just the ones 
that contribute the greatest amount to the overall
luminosity, so it would seem advisable to understand 
their dynamics as
thoroughly as possible.

In principle synchrotron light monitors could resolve 
changes in the
average beam profile and detect how they are modified by 
the collision process.
But the analysis of such measurements would be 
difficult because they are  
made far from the interaction point and depend on the 
highly 
distorted beta functions (at least for low amplitude 
particles)
as well as the beam distribution. 
Another technique that might address these types of 
beam-beam effects is
to shake one beam and 
measure the response of the
other \cite{SaganTechnique}. 
However while such measurements can provide zeroth order
and first order estimates for the beam density, there is 
no known method for
interpreting higher moments.

Given the luminosity as measured by the detector,  and 
the beam currents,
beamstrahlung measurements provide direct information 
about the average
vertical beam distribution during the collision. Like 
the other methods
mentioned above, interpretation is not obvious. By 
scanning the vertical offset
of the beams one should be able to see peaks in the 
beamstrahlung radiation
power, and a central minimum whose heights and depths 
depend on the
dynamically changing beam size. As mentioned earlier, 
polarization data from
the beamstrahlung might provide information about 
moments of the beam
distribution, particularly if there is a mismatch 
between the two beams.
These signals could be encorporated in feedback loops to 
aid in tuning and
improve the stability operation.

A proposed location for the beamstrahlung detector would 
be about 5 m from the
interaction point. Here there is essentially no hard 
sychrotron radiation from
the IR quadrupoles. About 2 W/cm of soft 
bend dipole radiation is deposited in this region with
a critical energy of 2 keV. This location is also 
favored by a
small beam stay-clear in both  planes 
which allows more penetration into the vacuum chamber 
with
mirrors. The average angle for beamstrahlung radiation 
striking the mirror
would be 6.2 mr. A mirror with a 1 cm penetration and 1 in 
diameter would
intercept a solid angle of $2\times 10^{-4}$ sr.

Collimation can be arranged so that the only  background 
synchrotron
radiation comes from the opposite side of the 
interaction point, which is about
10 m away, and such radiation is scattered once off the 
vacuum chamber.

\section{Backgrounds.}
We preliminarly study the CESR case, where the available 
beamstrahlung
power at the viewports is about 2 nW. As shown in 
Section~\ref{sn:Application}, 
the power deposited in the
region around the detector is about 2 W/cm along the 
main synchrotron
sweep. Therefore a background reduction factor of order 
$10^{-9}$ will suffice.
The backgrounds off the main sweep are perhaps three 
orders of magnitude less.
Three types of backgrounds are distinguished: radiation, 
visible and
irreducible backgrounds.

Radiation backgrounds are caused by radiation generating 
fluorescence
or Cherenkov light in the beamstrahlung detector itself.
Experience from Refs.~\cite{bonvi} and \cite{bonvi1} is 
of guidance. Radiation backgrounds drop very sharply 
with distance from
the beam pipe, thus only the radiation hitting the 
primary mirror and 
the primary window needs to be considered. Neither the 
window nor the mirror
are hit by the main sweep, and both could be somewhat 
recessed inside
the beam pipe.
The conversion factor for radiation energy into 
Cherenkov
light in glass-like media is about $10^{-4}$ for 
relativistic particles,
and virtually zero for synchrotron radiation (SR) with a 
critical energy of
less than 10 keV. The window would be well recessed from 
the beam pipe,
and shielded from beam halo, further reducing the 
Cherenkov backgrounds
by perhaps two orders of magnitude.

The conversion factor for fluorescence varies 
considerably
from material to material \cite{bonvi1}, but it can be 
conservatively estimated
to be $10^{-6}$ to $10^{-8}$.
For metals only, it is strongly dependent on the energy 
deposition mean depth,
as only the atoms at the skin depth can emit visible 
radiation.
The solid angle factor to a small iris could be of order 
$10^{-5}$ for both
window and mirror. By multiplying all 
the reduction factors together one obtains 
conservatively a reduction of
$10^{-11}$ for Cherenkov light in the window, and 
$10^{-14}$ for fluorescence.
These source of backgrounds should be 
negligible by several orders of magnitude.

Visible backgrounds (VB) are due to visible light 
emitted by the inner surface
of the beam pipe, either by beam pipe fluorescence, 
reflection and scattering
of visible light (mostly directly produced SR), or a 
combination of both.
At least at CESR, one detector would image the 
interaction point, and the
detector on the other side. Since the last scattering 
would happen on
the opposite detector, a variety of methods can be used 
to minimize VB.
Firstly, the flanges can be made somewhat rough, to 
insure that reflection
is isotropic and not forward peaked.
Most important of all, if the detector is truly 
symmetric in azimuth, 
individual viewports would be looking at another 
recessed viewport, if 
the detector is focussed on the other detector and not 
on the IP.
In all cases the solid angle factor is about $10^{-8}$, 
and the opposite detector is not hit by the main sweep. 
Thus a reduction
of at least $10^{-11}$ should be possible, and this 
source of backgrounds
is likely negligible.

Irreducible backgrounds are produced by particles in the 
beam tails, as they
get strongly focussed in the final quadrupoles. At CESR, 
there are perhaps
$10^{6}$ particles bent by more than 5 mrad in the 
vertical direction in the
last quadrupole. This kind of background is clearly 
irreducible, but falling
extremely sharply. A detailed calculation is complex, 
and beyond the scope
of this paper.
Should it prove to be a problem, moving the detection
frequency (and therefore the optimal angle) could prove 
to be the solution.

\section{Beamstrahlung Patterns: the FAQs}
In conclusion, we have shown that accurate beam-beam 
monitoring is
easily achievable at all existing or planned $e^+e^-$ 
particle factories.
The preferred method is the monitoring of large angle 
visible light 
beamstrahlung, which is produced in abundance 
(Table~\ref{tn:Machines}), 
and carries
information about the beam-beam collision conditions, 
not only limited
to luminosity but also beam-beam offsets, shape 
differences, and
other transverse mismatches.
The total diagnostic handbook goes far beyond the scope 
of a short paper,
but we wish to conclude with the Frequently Asked 
Questions file 
about beamstrahlung phenomenology \cite{gero}. Hopefully 
it might be of use
sometime in the future.

\begin{itemize}
\item {\it Q. How is red light polarization used?} 
Polarization is always
set to a minimum by initial steering, 
for optimal beam-beam overlap. The onset of
a non-minimal polarization during a run requires 
corrective action. Stable beam
conditions is obtained at a double local minimum point 
of minimal 
polarization and minimal power emitted. Zero 
polarization is achieved
only if the two beams match perfectly in the transverse 
plane.
\item {\it Q. How to do the steering to minimize 
polarization?} If the offset
is purely in the $x-$($y-$)axis, the counters at 90 and 
270 degrees will
show an excess in their respectively 
horizontal(vertical) polarization
counters. Remedial action is steering (say, for $y$ 
misalignments) up,
and if polarization increases, down. 
\item {\it Q. Can red light give information about 
optics optimization?}
Yes. Successfully squeezing the target beam will always 
result in higher power
from the radiating beam, and lower power for the target 
beam, for optimal
overlap. The reverse is also true.
As a general rule, in a comparison of equal energy, 
equally intense
beams, the wide beam radiates more than the narrow 
beam, all else being equal.
\item {\it Q. What if the two flat beams have a non-zero 
angle between
the major axes?} That will in principle generate a 
situation where the
minimal polarization is non-zero. The beam with the 
widest $y$ profile
(in case of flat beams, the one which has rotated) will 
radiate the most
in the vertical polarization, all else being equal.
\item {\it Q. 
What if the beams have different aspect ratios?} The 
rule of widest 
beam radiating the most is very far reaching. The widest 
beam in $y$ will 
radiate the most in $y$, the other beam will radiate the 
most in $x$, all else
being equal.
\item {\it Q. What can be learned from measuring the 
frequency spectrum
at large angle?} By moving, under stable beam 
conditions, 
the signal PMT across the visible spectrum, the beam 
length can
be measured accurately. This is a second reason to place 
the detector
at the optimal angle described above. A frequency scan 
at a substantially
lower angle would not work, because the gaussian factor 
becomes flat.
\end{itemize}

\end{document}